\documentclass[10pt]{article}
\usepackage{authblk}
\usepackage{amsmath,mathrsfs,graphicx,amssymb}
\usepackage{stmaryrd}
\newcommand{\ba}{\begin{eqnarray}}
\newcommand{\ea}{\end{eqnarray}}

\newcommand{\be}{\begin{equation}}
\newcommand{\en}{\end{equation}}

\newcommand{\bnu}{ \mbox{ \hspace{-1mm}\boldmath $\nu$ \hspace{-1mm}}}

\def\F{{\rm F}}
\def\K{{\rm K}}
\def\ev{\textbf{e}}

\def\mv{\textbf{m}}

\def\tv{\textbf{t}}
\def\rv{\textbf{r}}
\def\nv{\textbf{n}}

\def\d{{\rm d}}

\def\sp{{s^{\hspace{-0.5mm}+}}}
\def\sm{{s^{\hspace{-0.5mm}-}}}

\def\Lp{{L}}
\def\Lm{{l}}
\def\lco{{l_{\rm co}}}

\def\EE{{\rm E}}
\def\KK{{\rm K}}
\def\FF{{\rm F}}
\def\hyper{\;{\rm _p \FF_q}}

\def\km{k^{-}}
\def\kp{k^{+}}
\def\tf{\psi^-_f}

\def\pcr{P_{\rm cr}}
\def\lcr{l_{\rm cr}}

\def\lb{\llbracket}
\def\rb{\rrbracket}

\begin{document}

\title{Growth of a flexible fibre in a deformable ring}

\author[1]{\normalsize A. Cutolo}
\author[1]{\normalsize M. Fraldi}
\author[2]{\normalsize G. Napoli}
\author[3]{\normalsize G. Puglisi}

\affil[1]{\normalsize Dipartimento di Strutture per l'Ingegneria e l'Architettura, Università di Napoli Federico II, Italy}

%\affil[1]{\normalsize Dipartimento di Strutture per l'Ingegneria e l'Architettura, Università di Napoli Federico II, Italy}

\affil[2]{\normalsize Dipartimento di Matematica e Fisica “Ennio De Giorgi”, Universit\`a del Salento, Italy}

\affil[3]{\normalsize Dipartimento di Ingegneria Civile, Ambientale, del Territorio, Edile e di Chimica,
Politecnico di Bari, Italy}
\maketitle

\begin{abstract}
    We study the equilibrium configurations related to the growth of an elastic fibre in a confining flexible  ring. This system represents a paradigm for a variety of biological, medical, and engineering problems.  We consider a simplified geometry in which initially the container is a circular ring of radius $R$. Quasi-static growth is then studied by solving the equilibrium equations as the fibre length $l$ increases, starting from $l = 2R$. Considering both the fibre and the ring as inextensible and unshearable, we find that beyond a critical length, which depends on the relative bending stiffness, the fibre buckles. Furthermore, as the fibre grows further it folds, distorting the ring until it induces a break in mirror symmetry at $l>2 \pi R$.  We get that the equilibrium shapes depend only on two dimensionless parameters: the length ratio $\mu = l/R$ and the bending stiffnesses ratio $\kappa$. These findings are also supported by finite element simulation. Finally we experimentally validate the theoretical results showing a very good quantitative prediction of the observed buckling and folding regimes at variable geometrical parameters.
\end{abstract}

\section{Introduction}
\begin{figure}[t]
\centerline{\includegraphics[height=10 cm]{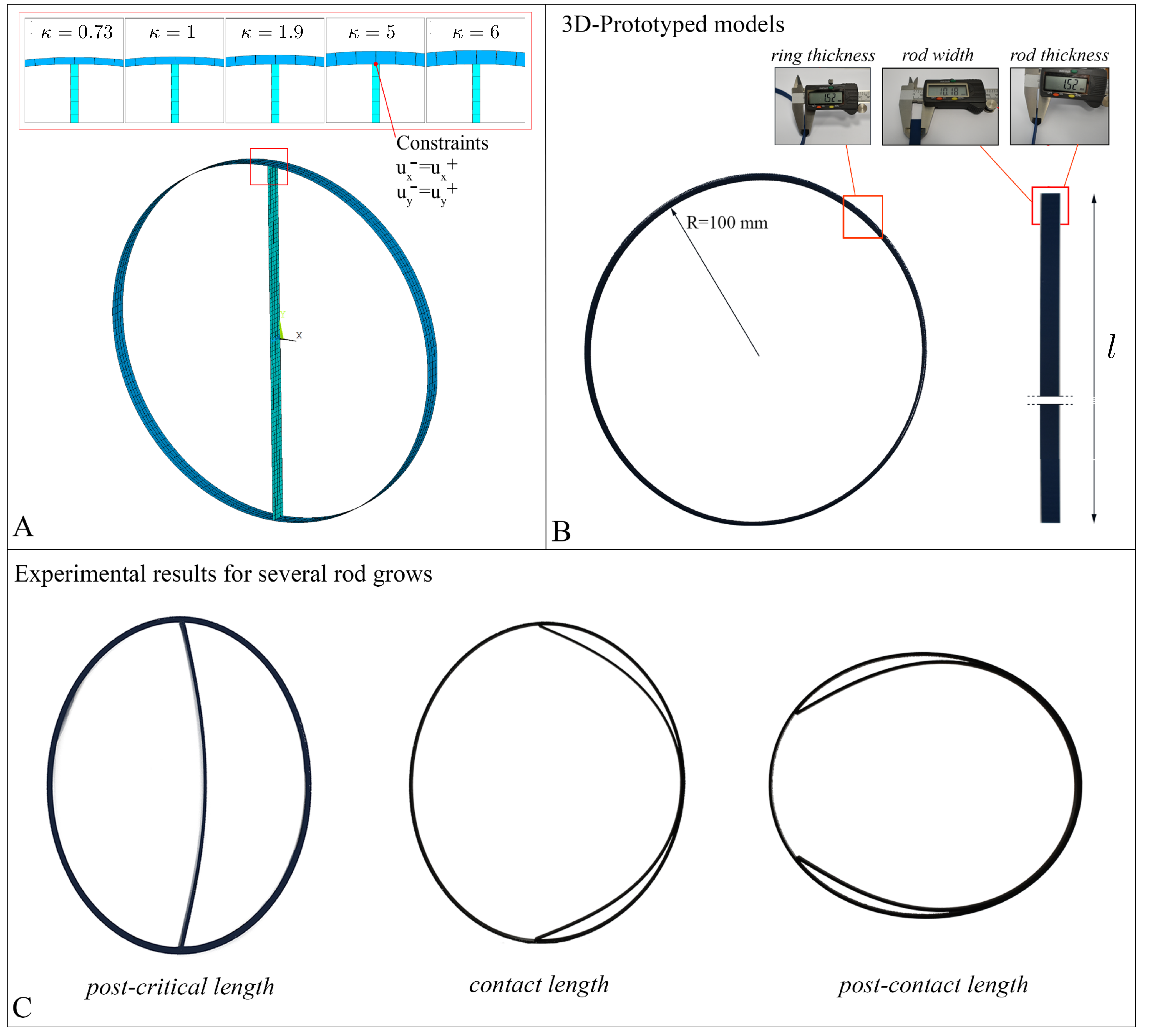}}
\caption{\label{figure:uno}Illustration of experimental settings. FEM scheme (panel A), design with 3D printer (panel B). Implementation of samples and three different shapes at varying fibre lengths (panel C).}
\end{figure}

To mimic typical confined growth phenomena, we consider the simple experiment sketched in Figure \ref{figure:uno}(c). We insert a fibre of length $\Lm$ into an elastic ring of length $\Lp$. As the length $\Lm$ exceeds the diameter of the ring, the end of the fibre unavoidably touch the surrounding wall. Then two cases can occur: i) the fibre remains straight by deforming the ring, or ii) the contact forces exerted by the ring at the ends of the fibre induce buckling of the fiber. The following questions naturally arise: What is the equilibrium shape as the fibre length  increases? How does it depend on the relative bending stiffnesses?

The system described above provides a paradigm for representing the shape and deformation of several physical systems at different length scales. Indeed, confined fibres or thin sheets  in  are  relevant for different technological applications,  ranging from packing steel plates into cylindrical containers for transport, to inserting flexible probes or catheters into body cavities for investigative or surgical reasons \cite{Schneider:2019}, and packing DNA into viral capsids \cite{Knobler:2009}. Consequently, studies on the mechanics, growth, and morphology of confined bodies are multiplying in various fields of Science \cite{Purohit:2003,Manning:2006, Lu:2008, Chen:2010, Stoop:2011,Kahraman:2012,Fosnaric:2013,Liu:2018, Elettro:2015, Elettro:2018, Li:2019,Trushko:2020,Croll:2022,Bartels:2022}. Furthermore, the analysis of  growth-induced instabilities represents an ideal playground to explore many systems of interest in engineering, biological, and medical applications \cite{Almet:2019, Su:2013}. These include cellular microtubules buckling induced by confinement of external microtubules networks in living cells  
by growth within vesicles \cite{Pallavicini:2017} or taking care of the role of surrounding citoplasma \cite{Jiang:2008}. This effect has been extensively explored in the framework of tensegrity cell models in \cite{Fraldi:2019}. An interesting and theoretical analysis of the influence of the traction of surrounding vesicles has been proposed in \cite{Elbaum:1996}. 
In addition, buckling effects of microtubules plays a key role also in regulating cell growth regulating plant morphology \cite{Bachmann:2019}. 

Previous studies have focussed on the packing of elastic rings, i.e. closed elastic rods, into undeformable circular cavities \cite{Cerda:2005, Boue:2006, DeTommasi:2021}, in the presence of capillary adhesion \cite{DePascalis:2014, Rim:2014} or friction \cite{Alben:2022}. Other studies have explored rods or strips in the presence of different types of confinement such as two parallel plates \cite{Elder:2020}, annular domains, \cite{Yang:2019} or polygonal confinement \cite{Hazel:2017}.
The possibility of deformable container is a generalisation of the above problems, in which the reciprocal interaction between the confined body and the container is taken into account. When both the confined body and the container are modelled as inextensible and unshearable flexible rods, the shape assumed by the system at equilibrium depends on  two only parameters: the ratio of lengths and the ratio of flexural stiffnesses \cite{Napoli:2017, Lombardo:2018}.

In some growth problems as well as in a cavity filament injection scheme, the confined body is a rod (with ends) instead of a loop \cite{Romero:2008, Gomes:2010, Gomes:2011, Wagner:2013, Napoli:2015, Elettro:2015, Napoli:2017a, Davidovitch:2021, Sobral:2021, Napoli:2022,Sherwood:2022, Yuan:2022}.  Under this circumstance, whenever the container is a ring, the equilibrium shapes and deformations have different features. A commonly observed feature is that, close to the edge, the rod
detaches from the wall and rejoins the cylinder only at the edge \cite{Romero:2008}.  This detached region is required since the fibre has to be straight at the point of contact. Such a result has been confirmed, both theoretically and experimentally, in a very recent paper \cite{Sherwood:2022}, where the force required to push the fibre into the container is also calculated.

In this paper, we consider the prototypical problem of confined growth-induced deformations and buckling of a system where both the fibre and the container are deformable (supposed inextensible and unshearable) with negligible friction. This study generalises previous results that consider a flexible rod confined by a rigid circular container. It also extends the results for loop-shaped confined bodies to fibres with ends. Analytical results are then confirmed  by finite thickness two-dimensional FEM analysis.   
 Eventually, we analyzed the effectiveness of the model in quantitatively describing the observed shapes by carrying out experimental investigations using elastic structures, subject to large deflections, manufactured with a three-dimensional printer. The growth mechanism of the rod is emulated by considering different lengths of the fiber, while the change in bending stiffness is obtained by considering several thicknesses of the fiber. The theoretical results are in a very satisfactory quantitative agreement with the experimental behaviour.
\section{Methods and material}
\subsection{Theoretical model}
Consider the planar equilibria of a one-dimensional elastic rod of length $\Lm$, which represents the fiber, naturally straight, constrained in a one-dimensional elastic ring of length $\Lp=2 \pi R$, that represents the container.
Fixing the length of the ring, we consider various lengths of the rod with its increase in length representing a growth process.

We can consider the growth process in two stages. The {\it early stage} starts when the length of the fibre exceeds the diameter of the ring ($\Lm> 2 R$) and ends at $\Lm = \lco$, when the midpoint of the fibre comes into contact with the ring. In this regime we assume that the only points of contact between the fibre and the surrounding ring are at the ends of the fiber. The {\it folding stage}, which begins when $\Lm$ exceeds $\lco$, is instead characterised by regions, in addition to the fibre ends, in which the ring and the fibre are in contact or the fibres intersects itself. The location and the length of these regions evolve as the relative length changes. 

Following the notation of \cite{Goriely:2017}, a point $\rv(s)$ on the  curve can be parametrised by its Cartesian coordinates
\be
\rv(s) = x(s)\ev_x +  y(s)\ev_y,
\label{kin1}
\en
where $s$ represents the arc length. Denoting by $\psi(s)$ the  angle between the tangent $\tv(s)$ and  the horizontal axis $\ev_x$, we have
\be
\tv(s)\equiv \rv'=\cos \psi(s)\ev_x + \sin \psi(s) \ev_y,
\label{kin2}
\en
where a prime denotes the differentiation with respect to $s$.  By replacing \eqref{kin1} into \eqref{kin2}, we get
\be
x'(s) = \cos \psi(s), \qquad y'(s)= \sin \psi(s),
\label{kin3}
\en
that have to be solved together with the mechanical balance equations given below.

Both the fibre and the ring must obey the balance of angular momentum that, in the absence of external distributed torques, reads \cite{Audoly:2010, Bigoni:2015, Goriely:2017}
\be
\mv'(s) + \tv(s) \times \nv(s) = {\bf 0},
\label{torque1}
\en 
where $\mv$ and $\nv$  are the resultant couple and force, respectively.  In addition, the balance of linear momentum  in the absence of external distributed loads assures that $\nv$  is constant along each edge. According to the Euler-Bernoulli theory of rods,  we assume that the internal bending moment for a planar deformation  is proportional to the curvature  $c(s)=   \psi'(s)$,  so that:
\begin{equation}
\mv(s) = k\, c(s) \ev_z,
\end{equation}
where $k$ is the bending stiffness and $ \ev_z= \ev_x\times \ev_y$.  Thus, \eqref{torque1} reduces to the second order ordinary differential equation
\be
k \psi''(s) - n_x(s) \sin \psi(s) + n_y(s) \cos \psi(s) =0,
\label{torque2}
\en
where $n_x$ and $n_y$, the horizontal and vertical components of $\nv$, are among the unknowns of the problem. As recalled above, by neglecting the effect of distributed loads (such as gravity and pressure)  $n_x$ and $n_y$ are constant.
\subsection{FEM simulation and experiments}
Both growth regimes described above were further analysed by means of numerical simulations and laboratory experiments. Firstly, a parametric finite element (FE) model was implemented, based on ad-hoc algorithms written in APDL (Ansys® Parametrical Design Language), to predict the effects of geometric and constitutive parameters on the equilibrium shape, including the progressive evolution of the contact regions after bifurcation. Since the structures considered were thin, both the outer ring and the inner fibre growth were carefully designed and discretized using shell elements (e.g. SHELL181), under the assumption of isotropic materials and finite displacements.

In addition, to validate the hypotheses and results of the theoretical and numerical modelling, several laboratory experiments were designed and performed under different conditions to faithfully replicate the different behaviours and mechanical configurations exhibited by theoretical results. To this end, equilibrium configurations, resulting from prescribed rod lengths associated with different relative bending stiffnesses, were selected, suitably designed, and then manufactured
through 3D printing additive manufacturing technologies. Specifically, all models were manufactured using ASA polymeric materials through a Stratasys F170 FDM printer.

\section{Early stage of growth}
\begin{figure}[t]
\centerline{\includegraphics[height=6 cm]{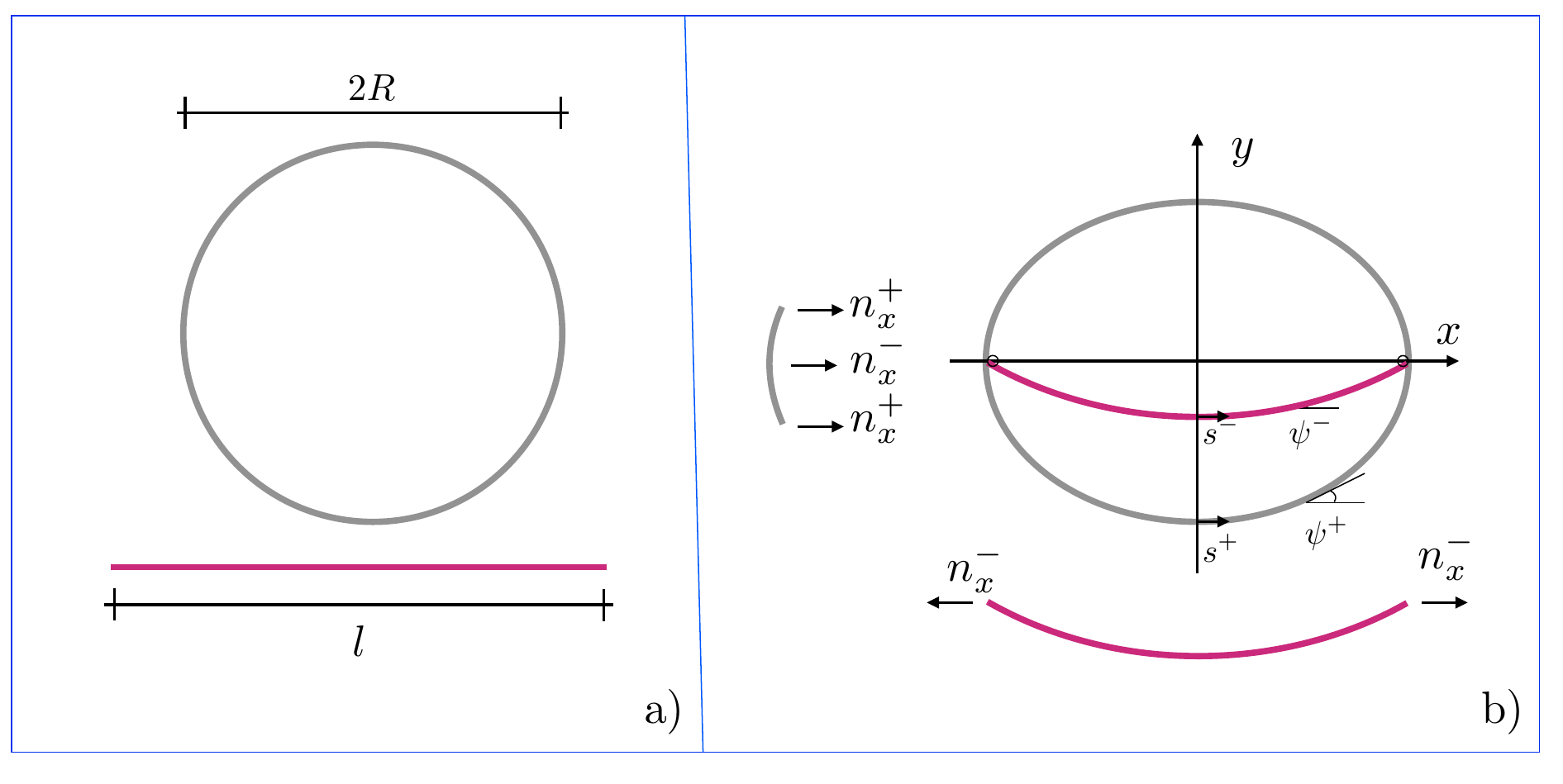}}
\caption{\label{figure:scheme} a) Undistorted fibre and ring. b) Buckling of the fibre due to the ring confinement.}
\end{figure}

We can assume that the ends of the fibre lie on the horizontal axis, the $x-$axis, and assume the origin of the axes at the midpoint of the two ends. Furthermore, let us assume that the fibre is mirror-symmetric with respect to the $y-$axis, while the loop is mirror-symmetric with respect to both the $x-$ and $y-$axes (see Figure \ref{figure:scheme}).

Here and in the following {\it the superscripts + and - refer to the ring and fiber, respectively}. Under the above assumptions, we can restrict our study to one half of the fibre $\sm \in [0,\Lm/2]$ and to one quarter of the ring $\sp \in [0,\Lp/4]$, where $\sm$ and $\sp$ denote the arc-length on the fibre and on the loop, respectively, as shown in Figure \ref{figure:scheme}.

\subsection{Growth-induced buckling}
Let us now address the following problem.  When $\Lm$ exceeds $2 R$ both the ends of the fibre unavoidably touch the loop. At the early stage the fibre remains straight while the ring deforms. Above a critical threshold length $\lcr$, the fibre is expected to buckle into a bent configuration. 
 
In order to asses $\lcr$, we assume that the fibre buckles symmetrically as  a {\it simple supported} rod subjected to an axial load. In this case, the classical analysis leads  to the formula for the Eulerian critical load of a beam with two hinged ends
\ba
\pcr := k_- \frac{\pi^2}{\lcr^2} = -n^-_x,
\label{pcr}
\ea
where $ k_-$ denotes the bending stiffness of the fiber. This formula is usually used to determine the critical load $\pcr$ of an elastic beam of a given length. However, unlike the classical problem, both $\pcr$ and $\lcr$ are unknown and are both affected by the properties of the surrounding loop.  Therefore, equation \eqref{pcr} must be coupled with the ring equilibrium problem that we will analyze below.

We notice that,  as a consequence of the mirror symmetry,  the vertical component of the internal force $n_y$ vanishes. This implies that equation \eqref{torque2} reduces to the nonlinear pendulum equation,
that  can be integrated leading to
\ba
\frac{1}{2}[\psi^+_{,\sp}(\sp)]^2 + h^+ \cos \psi^+(\sp) = c,
\label{eq:primo}
\ea 
where $h^+:= -n^+_x/\kp$ and $c$ is an integration constant. Then, by assuming that $\psi^+$ is an increasing function in $[0,\Lp/4]$, from \eqref{eq:primo} we get
\ba
\d \sp =  \frac{\d \psi^+}{\sqrt{2 c (1 - \eta \cos \psi^+)}},
\label{eq:ds}
\ea
where $\eta := h^+/c$,  that can integrated with the boundary conditions $\psi^+(0) = 0$ and $\psi^+(\Lp/4) = \pi/2$ to obtain 
\ba
\frac{	\Lp}{4} = \frac{1}{\sqrt{2c}}\int_0^{\frac{\pi}{2}} \frac{\d \psi^+}{\sqrt{1 - \eta \cos \psi^+}} =  \frac{1}{\sqrt{2c}}\left[\frac{\K(\xi)}{\sqrt{1-\eta}} + \frac{\eta}{2}  \hyper \left( \tfrac{3}{4},1,\tfrac{5}{4}; \tfrac{3}{2}, \tfrac{3}{2}, \eta^2\right)\right] 
\label{uno}
\ea
where  $\xi:= 2 \eta/(\eta -1)$, $\KK(\cdot)$ represents the complete elliptic integral of the first kind, and $\hyper(\cdot)$ denotes the generalized hypergeometric function \cite{Abramowitz:1970}. 

By using equation \eqref{kin3}$_1$ with the boundary condition $x^+(0) =0$, we get
\ba
x^+(\Lp/4) = \int_0^{\Lp/4} \cos \psi^+ \d \sp = \frac{1}{\sqrt{2c}}\int_0^{\frac{\pi}{2}} \frac{\cos \psi\, \d \psi}{\sqrt{1 - \eta \cos \psi}}
\ea
and, hence,
\ba
x^+(\Lp/4) =  \frac{1}{\eta \sqrt{2c(1-\eta)}}\left[(\eta -1) \EE(\xi) + \K(\xi) + \eta \sqrt{1-\eta} \hyper\left( \tfrac{1}{4},\tfrac{3}{4},1; \tfrac{1}{2}, \tfrac{3}{2}, \eta^2\right)\right],
\label{due}
\ea
where $\EE(\cdot)$  represents  the incomplete elliptic integral of the second kind \cite{Abramowitz:1970}.

We remark that, at the bifurcation threshold \eqref{pcr} we have $x^+(\Lp/4)=\lcr/2$, while the balance of force at the hinge connection with the rod (see the inset in Fig.~\ref{figure:scheme}) reads 
\ba
n_x^- =  -2 n_x^+.
\label{contact}
\ea 
that combined with equation \eqref{pcr} yields
\ba
\lcr = \frac{\pi}{\sqrt{2 \,c\, \kappa\, \eta}},
\label{lcr}
\ea
where $\kappa:=\kp/\km$ represents the relative bending stiffness. 

 Finally, by combining equation \eqref{due} and \eqref{lcr}, we obtain the  implicit equation for the sole unknown $\eta$: 
\ba
\frac{1}{\eta \sqrt{1- \eta}}\left[(\eta -1) \EE \left(\xi\right)  + \KK \left(\xi\right)\right] + \hyper\left( \tfrac{1}{4},\tfrac{3}{4},1; \tfrac{1}{2}, \tfrac{3}{2}, \eta^2\right) = \frac{\pi}{2 \sqrt{\kappa\, \eta}}
\label{eq:etacrit}
\ea
that can be solved numerically by varying the parameter $\kappa$.  Then $c$ can be obtained by using \eqref{uno}
\ba
c =  \frac{1}{\Lp^2 (1-\eta)}\left[\sqrt{2}\KK \left(\xi\right) + \eta \sqrt{1 -\eta} \hyper \left( \tfrac{3}{4},1, \tfrac{3}{4}; \tfrac{3}{2}, \tfrac{3}{2}, \eta^2\right) \right]^2; 
\ea
and finally $\lcr$ can be computed by means of \eqref{lcr}. 

\begin{figure}[t]
\centerline{\includegraphics[angle=0, width=1.0 \textwidth]{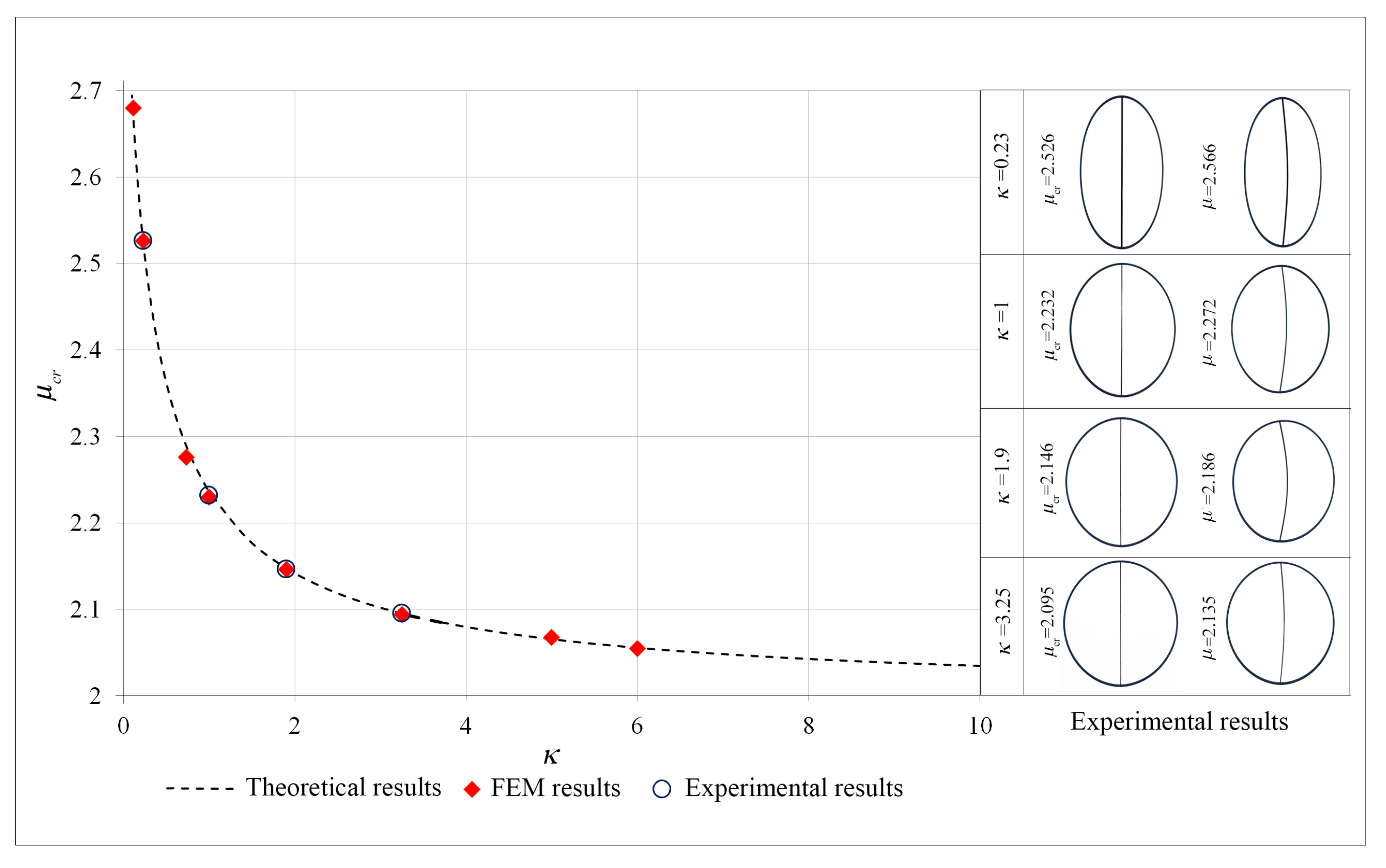}}
\caption{\label{fig:lcr} Relative dimensionaless critical length $\mu_{\rm cr}$ as a function of the relative stiffness $\kappa$. The dashed line corresponds to theoretical results, diamonds to FEM simulation and rings  to experiments, respectively.  The right-hand side of the figure illustrates the equilibrium shapes immediately below and immediately beyond the buckling threshold, for different values of $\kappa$.}
\end{figure}

Figure \ref{fig:lcr} sketches the dimensionless critical length $\mu_{\rm cr}:= l_{\rm cr}/R$ as a function of the ring-fibre relative stiffness $\kappa$. We observe that the critical threshold is a decreasing function of $\kappa$, so that the softer the disc, the longer the rod can grow without buckling. On the contrary, for large values of $\kappa$ (increasingly rigid containers), as expected the critical threshold tends to $2$, i.e. the length of the diameter of the circle of length $L$.

To validate the theoretical results, several samples were considered in which the dimensions of the outer ring are fixed (as shown in Figure 1), while fibres with different lengths and thicknesses were prototyped in order to reproduce certain values of the parameters $\kappa$ and $\mu$ used in the model.  Figure \ref{fig:lcr}   shows the comparison between our theoretical results and both  finite element simulations and experiments for different values of $\kappa$. In both cases, the agreement is excellent. The insets on the right side of Figure \ref{fig:lcr} sketch the experimental shapes for slightly shorter (left column) and slightly longer (right column) lengths of $\lcr$, for different values of $\kappa$. These
further support of the quantitative agreement between theoretical, FEM, and experimental behavior.
\subsection{Contact length}
%The above equations are supplied by suitable boundary conditions. Let us denote by $s^\pm=0$ the points on the y-axis, then left-right mirror symmetry implies
%\ba
%x^\pm(0) = 0, \qquad \psi^\pm(0) = 0,
%\label{bc1}
%\ea
%and the continuity conditions at the rod-ring contact point, where a $T-$node is formed,
%\begin{subequations}
%\ba
%x^-(\Lm/2) =  x^+(\Lp/4), \qquad  y^-(\Lm/2) =  y^+(\Lp/4) = 0
%\label{bc2}
%\ea
%\ba
%\psi^-_{,s^-}(\Lm/2) =0, \qquad \psi^+(\Lp/4) = \pi/2.  
%\label{bc3}
%\ea
%\ba
%n_x^- =  -2 n_x^+.
%\label{contact}
%\ea
%\end{subequations}
%These conditions apply when the length of the rod is between two critical quantities, the buckling length $\lcr$, at which the rod begins to bend, and the contact length $\lcon$, at which the midpoint of the rod contacts the ring. In the following we will show that both $\lcr$ and $\lcon$ depend only on the ratio between the bending constants of the rod and the ring.

Let us now calculate the length of the fiber, $\lco$, at which its midpoint touches the confining ring. This requires the integration of the equilibrium equation \eqref{torque2} for $s^- \in [0,\Lm/2]$, with the boundary conditions 
\ba
\psi^\pm(0) = 0, \qquad  \psi^-_{,s^-}(\Lm/2) =0,
\label{bc2}
\ea
expressing the mirror symmetry of the rod, and the condition that there are no torques applied to the ends of the rod,
respectively.

By combining  the boundary condition \eqref{bc2}$_1$ and the first integral of equation \eqref{torque2},  we get
\ba
\left[\psi^-_{,\sm}(\sm)\right]^2 = 2\, h^{-}\left[ \cos \psi^-(\sm)  - \cos \tf \right], 
\label{terza}
\ea 
where $h^{-}:= -n_x^-/\km$ and $\tf := \psi^-(\Lm/2)$. In the right-half portion of the rod, $\sm\in[0,\Lm/2]$, we assume $\psi^-(s)$ to be an increasing function, taking values between $0$ and $\tf$. Thus, equation \eqref{terza} can be solved by separation of variables
\ba
\int_0^{\Lm/2}\d \sm = \frac{1}{\sqrt{2 h^-}}  \int_0^{\tf} \frac{\d \psi^-}{\sqrt{\cos \psi^- - \cos \tf}},
\ea
yielding
\ba
\Lm =  \frac{ 2  \sqrt{2} {\F}(q_f)}{\sqrt{h^-(1 - \cos \tf)}},
\label{eq:elle}
\ea
where $q_f:= (\tf/2, \csc^2 \tf/2)$.

We now integrate equation \eqref{kin3}$_1$ for the rod so that
\ba
x^-(\Lm/2) =\int_0^{\Lm/2}\ \cos \psi^-(\sm) \d \sm = \frac{1}{\sqrt{2 h^-}}  \int_0^{\tf} \frac{\cos \psi^-  \d \psi^-}{\sqrt{ \cos \psi^- - \cos \tf}},
\label{e0}
\ea
whence we get
\ba
x^-(\Lm/2) =    \frac{\sqrt 2} {\sqrt{h^-(1  - \cos \tf)}}\left[ (1  - \cos \tf ) \EE(q_f) + \cos \tf {\rm F}(q_f)\right].
\label{eq:aa}
\ea
The contact between the rod end and the ring  implies $x^-(\Lm/2)=x^+(\Lp/4)$ that,  with the use of equations \eqref{due} and \eqref{eq:aa}, reduces to
\ba
\frac{1}{\eta \sqrt{2c(1-\eta)}}\left[(\eta -1) \EE(\xi) + \K(\xi) + \eta \sqrt{1-\eta} \hyper\left( \tfrac{1}{4},\tfrac{3}{4},1; \tfrac{1}{2}, \tfrac{3}{2}, \eta^2\right)\right] =\nonumber \\
=\frac{\sqrt 2} {\sqrt{h^-(1  - \cos \tf)}}\left[ (1  - \cos \tf ) \EE(q_f) + \cos \tf {\rm F}(q_f)\right].
\label{e2}
\ea

Now, we observe that in the correspondence of the contact points, the following boundary conditions hold 
\ba
y^-(0) = y^+(0), \qquad y^-(\Lm/2) = y^+(\Lp/4).
\label{bc4}
\ea 
Thus, by integrating equation \eqref{kin3}$_2$, we get
\begin{subequations}
\ba
y^-(\Lm/2) - y^-(0) = \int_0^{\Lm/2} \sin \psi^-(\sm) \d \sm = 
\frac{1}{\sqrt{2 h^-}}  \int_0^{\tf} \frac{\sin \psi^-  \d \psi^-}{\sqrt{ \cos \psi^- - \cos \tf}} = 
\sqrt \frac{2(1-\cos \tf)}{h^-},
\ea
\ba
y^+(\Lp/4) - y^+(0) = \int_0^{\Lp/4}\sin \psi^+(\sp) \d \sp  =
\frac{1}{\sqrt{2 c}}  \int_0^{\frac{\pi}{2}} \frac{\sin \psi^+  \d \psi^+}{\sqrt{ 1 - \eta \cos \psi^+}} =
\frac{\sqrt 2}{\eta \sqrt c} \left(1 - \sqrt{1-\eta}\right),
\ea
\end{subequations}
that replaced into \eqref{bc4} yields
\ba
\sqrt \frac{1-\cos \tf}{h^-}=\frac{1}{\eta \sqrt c} \left(1 - \sqrt{1-\eta}\right).
\label{e3}
\ea
Finally, we rewrite the condition \eqref{contact} as
\ba
h^- = 2 \kappa \eta \sqrt c.
\label{e4}
\ea
Equations \eqref{uno}, \eqref{e2}, \eqref{e3} and \eqref{e4} provide a system of coupled nonlinear algebraic equations for the unknowns $\eta$, $c$, $h^-$ and $\tf$. By replacing these values in \eqref{eq:elle} we obtain $\lco$.

\begin{figure}[t!]
\centerline{\includegraphics[angle=270, width=1.0 \textwidth]{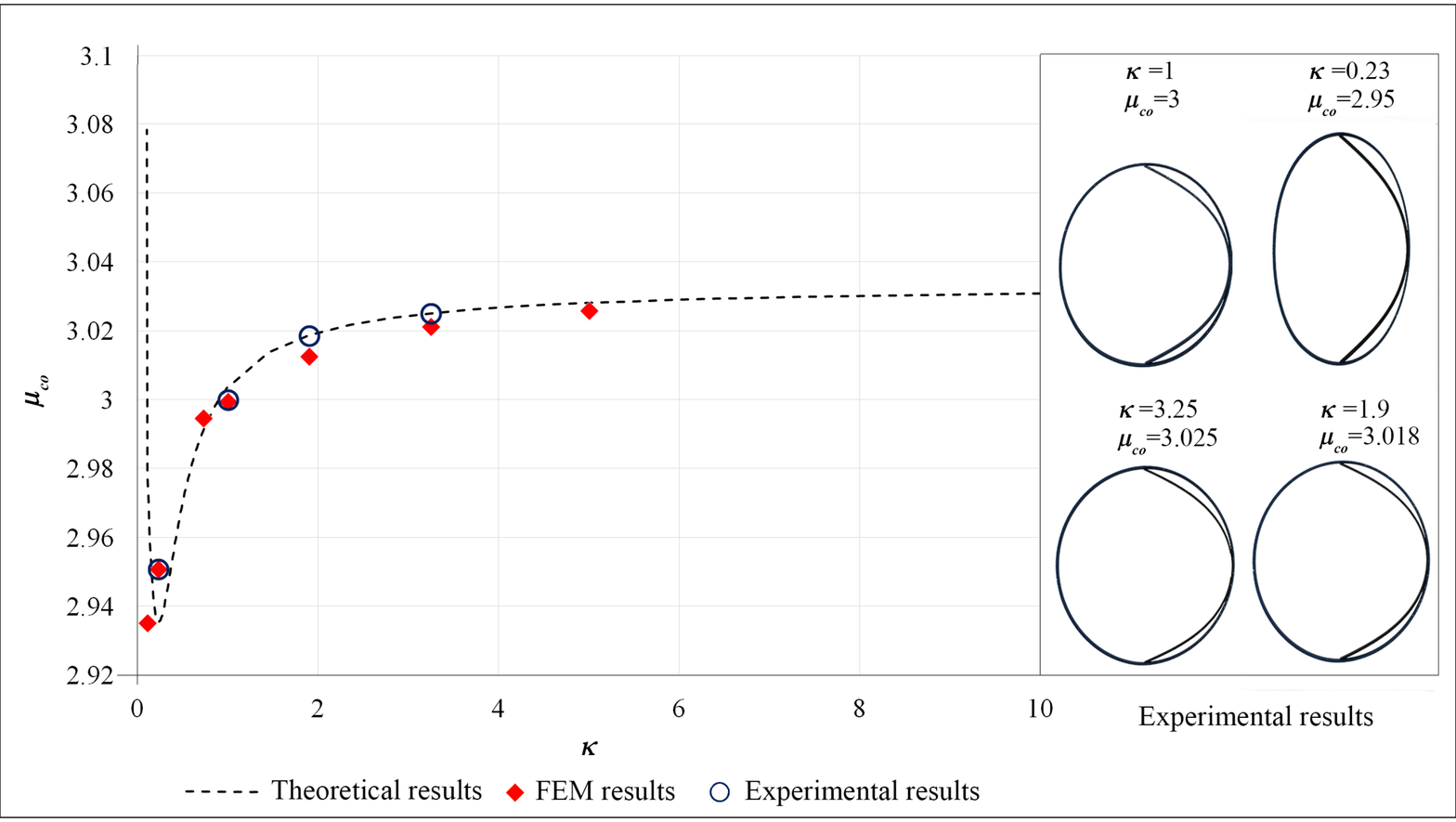}}
\caption{\label{fig:lco} Dimensionless contact length $\mu_{\rm co}$ as a function of the relative bending stiffness $\kappa$. The dashed line corresponds to theoretical results, diamonds to FEM simulation, rings  to experiments.  The right-hand side insets illustrate experimental equilibrium shapes at $l=\lco$, for different values of $\kappa$. }
\end{figure}

Figure \ref{fig:lco} shows the plot of $\mu_{\rm co} := \lco/R$ as function of the relative stiffness $\kappa$.  Again we attain a very good quantitative agreement of theoretical predictions with both experiments and finite element simulations.

\section{Growth beyond the contact length}
When $\Lm$ exceeds $\lco$, equilibrium forms are characterised by portions in which the ring and the fibre are in contact with each other or the fibre is in self-contact. These regions evolve as the relative length changes. Therefore, initially separate portions of curves may come into contact during the growth-induced packing process.  

\begin{figure}[t!]
\centerline{\includegraphics[angle=0, width=1 \textwidth]{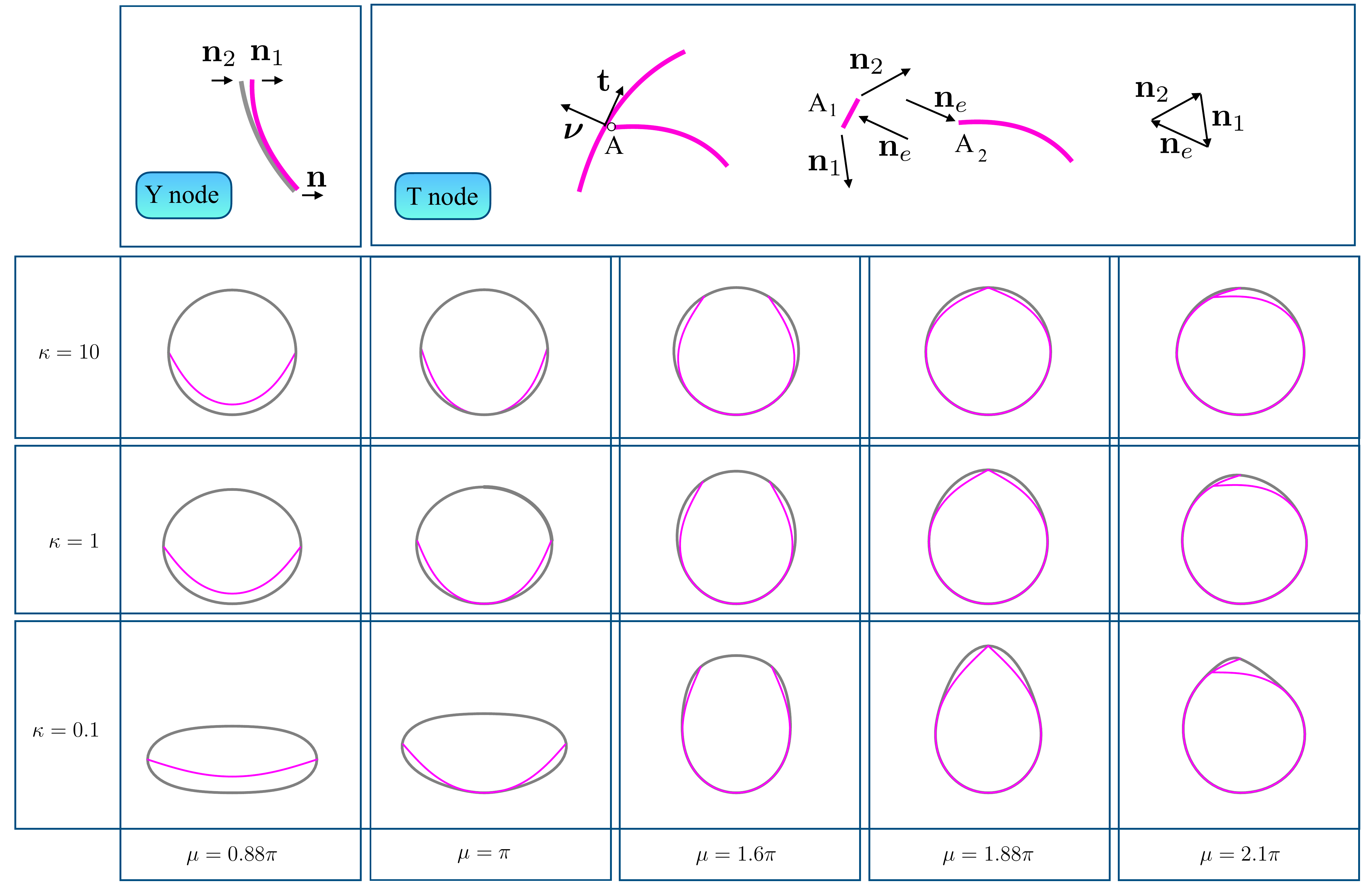}}
\caption{\label{fig:shapes} Theoretical equilibrium shapes as the length of the rod grows (from left to right) while the length of the outer loop is kept fixed. The different rows of shapes correspond to different amounts of relative bending stiffness. The container softens from the top to the bottom. Up insets represents the two types of nodes and corresponding jump equilibrium conditions.}
\end{figure}

According to one-dimensional modelling of the growth phenomenon, we neglect specimens thicknesses so that we may assume that when material points are in contact they share the same position in the plane. As a consequence, we may describe the portions of the system where curves are in frictionless contact as `effective' rods with bending stiffness coinciding with the sum of the stiffnesses of the constituent rods. In so doing, the structure can be understood as a closed graph with inextensible elastic edges of unknown length and different bending stiffness. Graph nodes (see Fig~\ref{fig:shapes}) are the points at which two separate curves branch off from two contacting curves ($Y$-node), or the points at which the ends of the confined rod touch parts of the ring or the rod itself ($T$-node).  Boundary conditions (see insets in Fig.~\ref{fig:shapes}) are imposed at the nodes whose locations are {\it a priori} unknowns.  

At a $Y$-node, a {\it mother edge} (the adhered region) bifurcates into two {\it daughter edges}. Following \cite{Napoli:2017}, force and curvature have to obey the equations 
\begin{subequations}
\be
\nv + \nv_1 + \nv_2 = {\bf 0},
\label{bilancioforze}
\en
\be
\psi' =  \psi'_1 =  \psi'_2,
\label{bilanciocurv}
\en
\end{subequations}
where  quantities without subscript refer to the  mother edge, while quantities with subscripts  1 and 2 refer to the daughter edges.  Furthermore,   \eqref{bilanciocurv} implies that  $x(s)$, $y(s)$, and $\psi(s)$ are continuous at the node, both along the fibre and along the ring.

At a $T$-node, an end of the rod pushes on a point on the ring or on the rod itself. The contact is assumed to be frictionless. Consequently,  denoting by $\nv_{\rm e}$ the force at the end of the fibre  and by $\nv$ the internal force of the portion with which it is in contact, with unit tangent $\tv$ and unit normal $\bnu$, we obtain 
\begin{subequations}
\be
(\lb \nv \rb - \nv_{\rm e} )\cdot \bnu = 0,
\en 
\be
\lb \nv \rb \cdot \tv =0, \qquad  \nv_{\rm e} \cdot \tv =0,
\en
\end{subequations}
where $\lb \cdot \rb$ denotes the jump of a quantity at the node for  the same portion of fibre or ring. In addition, since no torque is applied at the end of the fiber,  the boundary condition $\psi'_{\rm e}=0$ holds true.

\subsection{Results}
Using the Matlab routine bvp4c, we solve the equilibrium equations for different values of $\kappa$, keeping the length of the ring fixed and varying the length of the rod from slightly less than $\lco$ to slightly more than the length of the ring. Figure \ref{fig:shapes} illustrates the results  
as function of the  two dimensionless parameters $\mu$ and $\kappa$. 

For each $\kappa$ , it is possible to characterise common phases as the rod grows. At $\mu = \mu_{\rm co}$, the midpoint of the rod comes into contact with the ring. As the fibre grows further, this contact remains localized at a point, while the two end points move along the wall loop, until the rod length reaches a critical dimensionless length $\mu = \mu_1$. As this threshold is exceeded, the contact on the bottom spreads into an extended contact zone. This region increases in length with $\mu$, while the two end points of the fibre migrate towards the top of the ring, sliding on the walls.
When $\mu=\mu_2$, the two ends reach the axis of symmetry and come into contact with each other. The next growth phase develops with the ends in contact on the axis of symmetry until the length of the rod almost reaches that of the ring.

It is evident how relative stiffness affects equilibrium forms. In fact, Figure \ref{fig:shapes} shows that when the container is rigid $\kappa=10$ it remains almost circular throughout the growth process.  Conversely, when the ring softens ($\kappa=1$ or $\kappa=0.1$) it ovalises with the major axis changing direction during the different growth phases.

\begin{figure}[t]
\centerline{\includegraphics[angle=0, width=1 \textwidth]{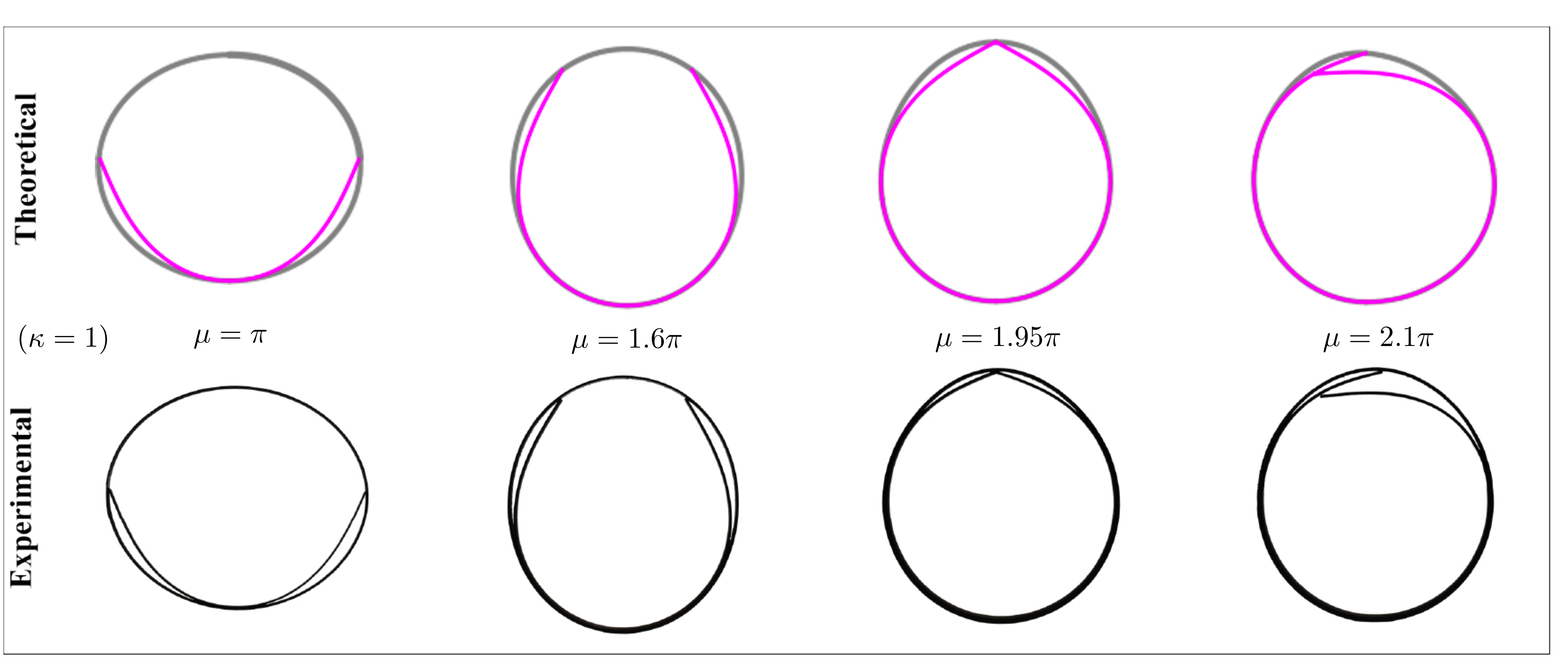}}
\caption{\label{fig:ab} Comparison between theoretical (top) and experimental (bottom) equilibrium shapes, for $\kappa=1$ and $\mu=\pi, 1.6 \pi,1.95 \pi,2.1\pi $. In experiments, an elastic strip of  length $l=334.00$ mm is inserted into four different rings of the same material and length $L= 215.20,134.04 ,109.18, 101.82$ mm. The width and the thickness for both ring and strip are $b=10.00$ mm and $h=1.24$ mm, respectively.}
\label{fig:ultima}
\end{figure}

{Eventually, Figure \ref{fig:ultima} shows the comparison between theoretical and experimental results for $\kappa=1$ at different lengths. In the experiment, both the fibre and the ring, of the same material, were designed with the same width and thickness, which ensures that $\kappa=1$. The different values of $\mu$ were obtained by considering progressively shorter ring lengths. Once again, the theoretical and experimental results are in excellent quantitative agreement.}

\section{Concluding remarks}
We presented a simplified model to describe how a flexible fibre grows and distorts a flexible ring surrounding it. Our analysis uses the classical Euler Elastica combined with non-trivial boundary conditions to describe the fiber-ring interplay. Elastica has the undeniable advantage of being integrable,  which allows us to obtain exact results. Our results show that at an early stage of the growth, when the fibre is slightly longer than the diameter of the ring, the fibre buckles after deforming the confining ring. Our analysis determines this critical length as a function of the relative bending stiffness. As the fibre grows further, we see it packing, leading to a distortion of the container that is more pronounced when the confining loop is softer. Several theoretically obtained shapes of equilibrium are confirmed by experiments.

Our results could be an inspiration to extend some recent findings to the case of soft containers. For example, an extension of the results in \cite{Sherwood:2022}, could be obtained by considering appropriate boundary conditions to describe the direction of wire injection into the cavity. Another significant effect to consider is the presence of friction, which would increase the class of equilibrium configurations, with the presence of multiple solutions, and introduce new effects such as rolling contact and hysteresis \cite{Alben:2022}. Finally, we mention a possible extension to lengths far greater than those we have considered, as in  \cite{Romero:2008}, to determine how a multiple coiled elastic sheet is configured in an elastic tube. Furthermore, our results indicate that, behind the theoretical interest of the proposed model, it can constitute an inspiration in the field of several actuation and sensing system, e.g. in the field of NEMS and MEMS.

%\enlargethispage{20pt}

%\ethics{Insert ethics text here.}

%\dataccess{All supporting data are provided within the paper.}

%\aucontribute{All the authors equally participated in the design of the research, in developing the analysis, in the discussion and interpretation of the results, prepared the figures and wrote the manuscript.}

%\competing{We declare we have no competing interests.}

\section*{Acknowledgements} 
{A.C. and M.F. acknowledge the support of MIUR (Italian Ministry of Education, University and Research)  through the grant PRIN-20177TTP3S. The work of GN. has been funded by the MIUR  Project PRIN 2020, “Mathematics for Industry 4.0”, Project No. 2020F3NCPX.  G.P. has been supported by the MIUR-PRIN project 2017KL4EF3. This manuscript was also conducted under the auspices of the GNFM-INdAM.}

%\ack{Insert acknowledgment text here.}

%\disclaimer{Insert disclaimer text here.}

%\bibliographystyle{unsrt}
%\bibliography{sudici}

\end{document}